\documentclass[12pt,aps,prd,preprint,showkeys]{revtex4}
\usepackage{graphicx}

\usepackage{xcolor}
\usepackage{subcaption}
\usepackage{ulem}
\usepackage{csquotes}
\usepackage{subcaption}
\usepackage[percent]{overpic}

\date{}

\begin{document}

\title{Classification of flavor dependence of Chiral Magnetic Effect with Deep Neural Network using multiple correlators.}
	\author{Somdeep Dey$^1$}
	\author{Abhisek Saha$^{2,3}$}
	\author{Soma Sanyal$^1$}
	
	\affiliation{$^1$ School of Physics, University of Hyderabad, Gachibowli, Hyderabad, India 500046}

	\affiliation{$^2$ School of Physics, Peking University, Beijing 100871, China}
	\affiliation{$^3$ Center for High Energy Physics, Peking University, Beijing 100871, China}

\begin{abstract}
We study the flavor dependence of the Chiral Magnetic Effect (CME) by analyzing two key charge-separation correlators used to characterize the charge separation effect: the conventional $\Delta\gamma$ and the recently proposed $R_{\psi_2}$. Using the AMPT (A Multiphase Transport) model with an initial-state centrality-dependent charge separation, we evaluate the sensitivity of these correlators to 2-flavor ($u,d$) and 3-flavor ($u,d,s$) quark scenarios. While both correlators exhibit modest flavor dependence in mid-central (30–50\%) collisions, their discriminative power varies significantly with centrality and transverse momentum ($p_T$), limiting their utility disentangling the flavor dependent scenarios.
To overcome these limitations, we develop a neural network classifier trained on final-state hadronic observables (e.g., $dN_{ch}/d\eta$, $p_T$ spectra). The model achieves $>90\%$ accuracy in flavor classification by leveraging multi-observable correlations, with $p_T$-differential features proving particularly discriminative. Crucially, by incorporating background contributions directly into the training data, our approach provides more reliable flavor estimates than correlator-only methods.

\end{abstract}


\maketitle
\section{Introduction} 			\label{sec:intro}

The collisions at the Relativistic Heavy Ion Collider (RHIC) and the Large Hadron Collider (LHC) result in the formation of the quark gluon plasma (QGP). The formation of the QGP indicates a phase transition from the hadronic to the QGP phase. The phase transition based on Quantum Chromodynamics (QCD), can result in metastable domains with variation in the topological charges of the gluonic fields\cite{domain1,cme_exp}. The quark interactions with these fields can result in a chirality imbalance which in the presence of a strong magnetic field can result in the separation of the positive and negative charges along the magnetic field direction. This is referred to as the "Chiral Magnetic Effect" (CME) and it's importance lies in the fact that it can be seen as a direct experimental signature of the topological fluctuations of the QGP \cite{CME,CME1,cme1,cme2,cme3}.       

The challenge in looking for the CME signal comes from the fact that charge separation in heavy ion collisions (HICs) occurs due to various different physical processes. Though a charge separation has been reported by the STAR Collaboration \cite{STAR1, STAR2}, the separation may be due to resonance decays or charge ordering in jets etc. Moreover, since a CME signal means that the charge separation must originate from domains of charged particles with the same parity, the charge correlator has to be sensitive to the parity of the charged particles too\cite{parity}. Hence specialized charge correlators have to be defined which take into account all these factors. The standard correlator used to measure the CME in heavy ion collisions is the $\gamma$ correlator \cite{gamma1}. This is a charge dependent azimuthal correlator given by, 
\begin{equation}
    \gamma_{\alpha\beta} = \langle cos(\phi_\alpha + \phi_{\beta} - 2 \Psi_{RP})\rangle
    \label{eq:gamma}
\end{equation}
where $\phi_{\alpha}$ and $\phi_{\beta}$ are the azimuthal angles of two charged particles $\alpha$ and $\beta$ respectively and $\Psi_{RP}$ is the reaction plane angle. $\alpha$ and $\beta$ can be either same charge(SS) or opposite chage (OS). This correlator has been investigated both theoretically as well as experimentally \cite{Huang1,Zhao,gamma2,cme_exp}.
  
The phase transition dynamics of Quantum Chromodynamics is still not well understood. The critical temperature for the phase transition as studied in lattice QCD gives different values for the two flavor and three flavor QCD \cite{tc_flv,tc_flv1,tc_flv2}. For the Chiral effects too, the number of flavors are important. The electric current induced along the external magnetic field is given by, 
\begin{equation}
\vec{J} = N_c \left( \sum_f Q_f^2 \right) \frac{e^2}{2 \pi^2} \mu_5 \vec{B}
\end{equation}  
where $ N_c = 3 $, is the number of color and the sum over $N_f $ is over the flavors with electric charge $ Q_f $ respectively \cite{CME,j_mu1,j_mu2}. Therefore the electric current will vary for two flavor and three flavor cases. The quark flavor dependence of the CME has been investigated by Huang et. al. \cite{Huang1} for Pb+Pb collisions using the $ \gamma $ correlator mentioned before. They have used the AMPT model, modified to involve CME in the initial condition and have obtained the $ \gamma $ correlator for the two flavor and the three flavor cases separately. They have found that though there are differences in the values of the $ \gamma $ correlator for the two cases, the overall statistics were not enough to disentangle the two scenarios unambigously. 

In this work, we use a different method to introduce the CME in the AMPT model. The method we use has been previously used to introduce CME in the PACIAE model \cite{paciae}. We first check the  $ \gamma $ correlator
for the two flavor and three flavor cases using this method. We find that the results obtained are similar to the ones obtained by Huang et. al. \cite{Huang1}. Though, a fine difference is observed in the $\gamma$ correlator for 2-flavor and 3-flavor cases, the difference is not very clear and depends on the centrality and $p_T$ ranges. We improve on these studies by creating a classification model with a neural network architecture and train the model using numerous combinations of the final state particle distributions. We also determine which observables are best suited for precise and accurate CME flavor estimation.

Apart from the $ \gamma $ correlator, the other correlator which is used to detect the chiral magnetic effect is the correlation function $ R_{\psi_m} (\Delta S)$ \cite{magdy1}. This correlation function depends on the angles between the different event planes generated from the Fourier decomposition of the charged particle azimuthal distribution. The reaction plane is generally defined by the plane defined by the impact parameter and the beam axis. This is referred to as $\psi_{RP}$. The Fourier decomposition of the charged particle azimuthal distribution is given by 
\begin{equation}
\frac{dN^{ch}}{d\phi} \propto 1 + 2 \sum_n (v_n cos (n\Delta \phi) + a_n sin (n\Delta \phi) + ....)
\end{equation}
where $\Delta \phi = \phi -\psi_{RP} $. This basically gives the particle azimuthal angle with respect to the reaction plane angle. The coefficients of the P- even and P-odd Fourier terms are given by $v_n$ and $a_n$. The correlation function  we use is related to the correlation with the $\psi_2$ plane and is refered to as the $ R_{\psi_2} (\Delta S)$. This second correlator was suggested as the $\gamma$ correlator is challenging to measure for the very small CME induced charge separation and it can be contaminated with background effects that mimic the charge separation. The charge separation is inherently very small as the the number of correlated particles from the same P - odd domains are small. So after the expansion and the hadronization the net charge separation due to CME would be difficult to identify amongst the other background fluctuations. The $ R_{\psi_2} (\Delta S)$ correlator is defined by, 
\begin{equation}
R_{\psi_2} (\Delta S) = \frac{C_{\psi_2}(\Delta S)}{C_{\psi_2}^{\perp}(\Delta S)}
\end{equation}
where $ C_{\psi_2}(\Delta S) $ and $ C_{\psi_2}^{\perp}(\Delta S) $  are correlation functions to quantify the charge separations  parallel and perpendicular to the magnetic field. Since the CME related charge separation occurs along the direction of the magnetic field and $\psi_2$ and $ \psi_{RP} $ are strongly correlated, the numerator measures both the CME and the background driven charge separation. However, the denominator which determines the charge separation perpendicular to the magnetic field is determined only by the background driven charge separation. So the ratio of the two would give us an indicator of the CME driven charge separation. We will later on mention in detail how these quantities are calculated in our work. 

The important point that we aim to address is whether the flavor dependence seen for the $\gamma$ correlator is also reflected in the other correlator. Based on the charge separation, it has been found that the $ R_{\psi_2} (\Delta S)$ correlator gives a concave distribution whose width $a_1$ is proportional to the initial charge separation parameter. Basically, the width of the concave shaped distribution determines the magnitude of the charge separation in this case. A narrower width indicates a higher charge separation compared to a broad width concave distribution.  

In the case of the $\gamma$ correlator, the charge separation is found to be different for the two flavor and the three flavor cases, the three flavor case showing a larger value of the $\gamma$ correlator as compared to the two flavor case. We can draw a similar conclusion for the  $ R_{\psi_2} (\Delta S)$ too, we find that the width of the concave distribution is different for the two flavor and three flavor cases. The three flavor case shows a reduced width compared to the two flavor case. A narrower concave shaped distribution indicates a stronger CME driven charge separation. So in essence, both the correlators indicate similar quark flavor dependence of the chiral magnetic effect. This means that with proper experimental data after taking care of the backgrounds, we should be able to distinguish the flavor dependency of the CME. 

The numerical computation of these correlators however are computationally challenging. They require large number of ensemble averaging to achieve results that can be matched with the experimental data. Especially for the $R_{\psi_2} (\Delta S)$ correlator which requires averaging over two directions.  
Smaller number of ensembles give larger errors and the distributions cease to be distinctive. Since our motivation is to identify the two flavor and three flavor CME, we felt that it should be possible to use 
Machine learning (ML) techniques for classification so that the computational challenges are reduced.  
Based on this, we develop a binary classification model based on the Deep Neural Network architecture to classify the CME into two groups based on the number of flavors. We choose the Convolutional Neural Network (CNN) as the basis of our model as it is known to out-perform other models in classification accuracy. The CNN would help us in streamlining the classification process while maintaining high accuracy. CNN has been used previously to model the data in relativistic heavy ion collisions \cite{DNN}. Here we have used it to build a surrogate model that speeds up the predictions without sacrificing too much accuracy.

In section II, we discuss the introduction of CME into the AMPT model. We also define the various correlators in more detail based on the model. In section III, we share the general results obtained for flavor dependence for the $\gamma$ correlators and in section IV, results for flavor dependence of the $R_{\psi_2} (\Delta S)$ are presented. Section V discusses the classification based neural network model that we train to understand the important features of the flavor dependence of the CME. Finally in section VI, we summarize our results and discuss the conclusions obtained from this work.

\section{The AMPT model} 			
\label{}
One of the most well established model to study the different aspects of relativistic heavy ion collision is the AMPT model \cite{ampt,ampt2}. This model is a hybrid transport model based on the Monte Carlo algorithm. It has four main stages. The first stage is the initial stage. The initial condition, including the spatial and momentum distributions of minijet partons and soft string excitations, is obtained from HIJING model \cite{hijing,hijing1,hijing2}. This is followed by the parton cascade, the hadronization and the final hadronic scattering stage. Details of this model is available in several references \cite{amptref1,amptref2}. In our calculation, the string melting version of AMPT model is used to simulate the HIC evolution. The direction
of magnetic field is on -y direction on average. By default there is no CME mechanism in the current AMPT model. Generally, the initial charge separation is introduced by exchanging the $p_y$ values of a certain percentage of downward moving quarks with those of the same number of upward moving antiquarks of the same flavor. The exchange has to be done in a way that the total momentum is conserved. This method was first used in ref. \cite{Ma} and has subsequently been used in other studies. The switching percentage is defined by, 
\begin{equation}
f = \frac{(N^{+ (-)}_{\uparrow(\downarrow)} - N^{+ (-)}_{\downarrow(\uparrow)})}{(N^{+ (-)}_{\uparrow(\downarrow)} + N^{+ (-)}_{\downarrow(\uparrow)})}
\end{equation}
where + and - indicate the quarks' positive and negative charges, N is the number of quarks of a certain species (u, d, or s), and $\uparrow$ and $\downarrow$ indicate whether the quarks are moving parallel to or anti-parallel to the magnetic field.
Another method of introducing the CME is using the method which was used in the simulation model PACIAE \cite{paciae}. In PACIAE model, the switching fraction is given by a semi empirical function. This semi empirical function is obtained from experimental results and the basics of collision dynamics in an electromagnetic field. Considering $A$ to be the atomic mass number of the colliding nucleus and $Z$ to be the atomic number, $b$ the impact parameter and $\sqrt{s}$ the center of mass energy, we have,
\begin{equation}
f \propto  C  \left(C_{1} b A^{-5/3}+ C_2 b^3 A^{-7/3} \right) \left( C_3 + C_4 \frac{\sqrt{s}}{E_{RHIC}} {(\frac{Z}{A})}^{2/3} \right)
\label{eq:f}
\end{equation}   
Here $E_{RHIC}$ is the RHIC energy i.e., $\sqrt{s_{NN}}=200$ GeV and the constants $C, C_1, C_2, C_3, C_4 $ are obtained by fitting empirical polynomial curves to the electromagnetic fields in the experimental set up.  In this work, we have used this empirical method of determining the value of the switching fraction $f$. So the CME is introduced in the initial stage of the AMPT for a particular value of $f$ and the final data is analysed to check for signatures of two or three flavor CME. Further we imposed a condition to select quarks based on the flavors. For a 2 flavor scenario, the initial momentum exchange was incorporated in $u$ and $d$ quarks specifically and in 3 flavor case, this has been extended up to the strange quarks.

\begin{figure}[h!]
    \centering
    \includegraphics[width=0.5\linewidth]{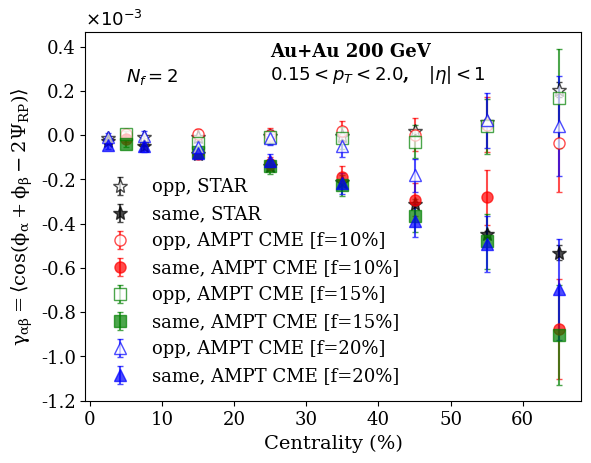}
    \caption{Plot of $\gamma$ correlation vs centrality for same and opposite charged particles in Au+Au collision at $\sqrt{s_{NN}}=200$ GeV for $N_f=2$ with different charge separation fractions ($f$), along with the STAR data. }
    \label{fig:gammaC1}
\end{figure}

First we analyze the gamma correlator for different initial conditions of the collision system. In Fig. \ref{fig:gammaC1}, we show the dependence of $\gamma$ correlator on the charge separation fraction for Au+Au collision at a center of mass collision energy ($\sqrt{s}$) of 200 GeV. We generate $5\times 10^5$ number of minimum bias Au+Au collision events and the centrality is determined by the number of charged particles at mid-rapidity ($\frac{dN_{ch}}{d\eta}_{|\eta|<1}$). The $\gamma$ correlation is obtained from the Eq. \ref{eq:gamma} and $f$ is calculated using Eq. \ref{eq:f}. The reaction plane is approximaed as the event plane angle and reconstructed from the azimuthal distribution of particles produced  at midrapidity:
\begin{equation}
    \Psi_{EP} = \frac{1}{2}tan^{-1}\left[ \frac{\sum w_i \sin(2\phi_i)}{\sum w_i \cos(2\phi_i)}  \right],
\end{equation}
where $w_i$ is a weight for each particle \cite{psi_ep}. In our calculation, we use particle transverse momentum $p_T$ as weights.
Different percentages of charge fraction denoted by $f$ are shown by different colors and shapes in the plot. We have also plotted the experimentally obtained $\gamma$ correlator from the STAR collaboration \cite{STAR1} for Au+Au collision at $200$ GeV shown by the black star symbols. We find good agreement with the STAR data which validates our introduction of CME in the initial stage of the AMPT code. The correlation is very weak in the central collisions and becomes larger in the peripheral collisions\cite{peripheral_mag}. This may have been caused by the dilution of correlation which occurs in the case of particle production from multiple sources. This can be compensated by multiplying the number of nucleon participants. Also we expect smaller correlations in most central collisions as weaker parity violation is expected as the magnetic field weakens. Opposite-charge correlations are very
 small compared to the same-charge correlations. This is because of the suppression of the back-to-back correlations and it is also seen in the experimental data. For larger charge separation fraction, the $\gamma$ correlation increases and the increase is more for large centrality .i.e., for peripheral collisions. We found good agreement up to $50-60\%$ centrality. After that our model overpredicts the experimental data. 

\section{Flavor dependency in $\gamma$ correlator}
\begin{figure}
    \centering
    \begin{subfigure}[b]{0.45\textwidth}
        \begin{overpic}[width=\textwidth]{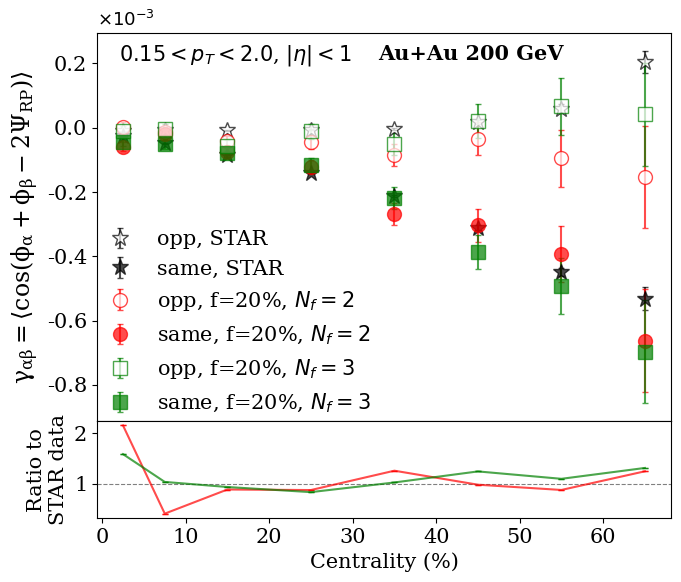}
            \put(18,55){(a)} 
        \end{overpic}
    \end{subfigure}
    \hfill
    \begin{subfigure}[b]{0.45\textwidth}
        \begin{overpic}[width=\textwidth]{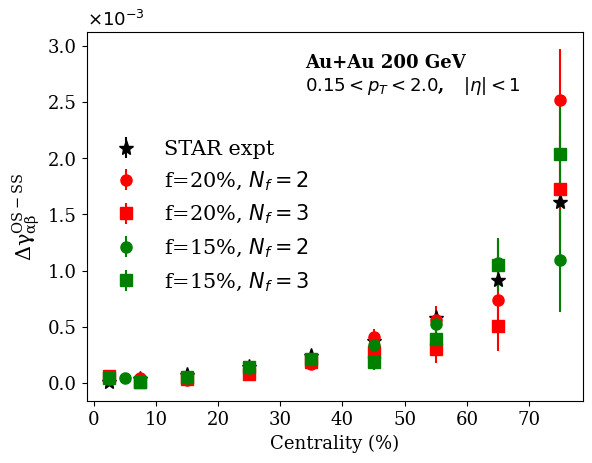}
            \put(20,65){(b)}
        \end{overpic}
    \end{subfigure}
    \caption{(a) Plot of $\gamma$ correlation vs centrality($\%$) for same and opposite charged particles in Au+Au collision at $\sqrt{s_{NN}}=200$ GeV for $N_f=2$ and $N_f=3$ with $f = 20\%$, with the STAR data. The bottom panel shows the comparison of the ratio of the individual cases with the STAR data (b) Plot of $\Delta \gamma^{os-ss}_{\alpha\beta}$ for $N_f = 2$ and $ N_f = 3 $ for different  $f\%$ along with the STAR data.}
    \label{fig:gammac3c4}
\end{figure}
In Fig. \ref{fig:gammac3c4}(a), we compare the $\gamma$ correlator for 2-flavor and 3-flavor scenarios, assuming an initial charge separation fraction of $f=20\%$. The results are plotted as a function of centrality for Au+Au collisions at $\sqrt{s_{NN}} = 200$ GeV, alongside the STAR experimental data \cite{STAR2} .
Here we find that the $\gamma$ correlator values for both 2-flavor and 3-flavor exhibit minimal separation, with large statistical uncertainties at higher centralities beyond $50\%$. Both scenarios agree reasonably well with the STAR data within errors, making it challenging to differentiate which flavor dynamics dominate the experimental signal.
The ratio plot (Fig. \ref{fig:gammac3c4}, bottom panel) quantifies the agreement between theory and experiment by normalizing the model predictions to the STAR data \cite{STAR2}. Notably, For central to mid-central collisions (0–50\%), the 3-flavor ratio hovers near unity, reinforcing its consistency with data. The 2-flavor scenario shows slight deviations (5–10\%), particularly in the 30–40\% centrality bin. The rise in the ratio for peripheral collisions ($>60\%$) further underscores the breakdown of both the CME signal in this regime, possibly due to dominant background processes or dilution of the magnetic field. 
The closer agreement of both 2-flavor and the 3-flavor model suggests that strange quarks ($ss^{-}$) may play a non-negligible role in the CME-driven charge separation, as their larger mass and delayed thermalization could affect parity-violating currents \cite{parity}.\\

To isolate the CME-sensitive component, Fig. \ref{fig:gammac3c4}(b) shows the difference $\Delta \gamma(=\gamma_{OS}-\gamma_{SS})$ between opposite-sign ($\gamma_{OS}$) and same-sign ($\gamma_{SS}$) correlators. This observable cancels charge-independent backgrounds (e.g., local momentum conservation\cite{pt_cons,pt_cons1}) and highlights the CME-driven charge separation. Here, we find a clear distinction between the 2-flavor and the 3-flavor predictions, with the 2-flavor scenario (green circles) aligning more closely with the STAR data up to 50-60\% centrality. Both the CME signals overpredict $\Delta \gamma$ for peripheral collisions ($>70\%$), suggesting either non-CME contributions or limitations in the assumed initial charge separation. The overprediction in peripheral collisions can also be caused due to the decreasing lifetime of the QGP and the magnetic field \cite{peripheral_mag,peripheral_mag1}, reducing the CME signal relative to backgrounds like resonance decays or jet fragmentation. The large errors at high centrality highlight the need for higher-statistics data or improved background subtraction methods to constrain CME signatures definitively.\\

\begin{figure}
    \centering
    \begin{subfigure}[b]{0.45\textwidth}
        \begin{overpic}[width=\textwidth]{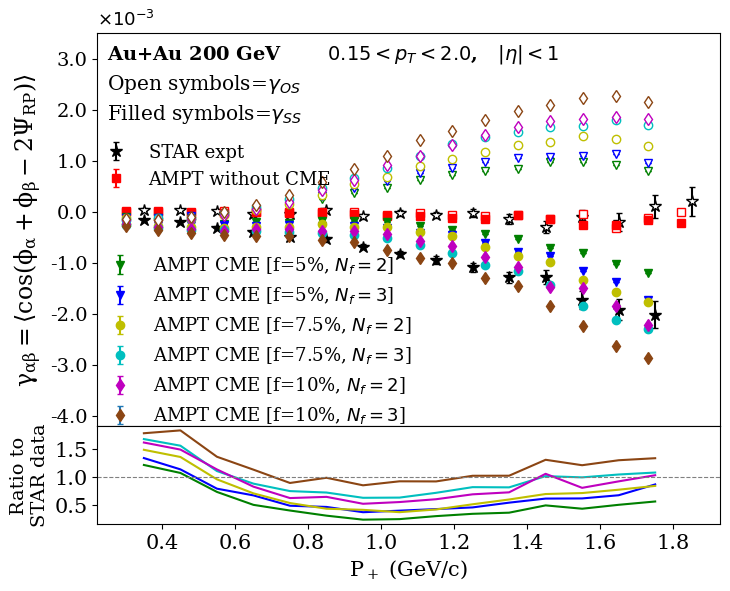}
            \put(58,25){(a)} 
        \end{overpic}
    \end{subfigure}
    \hfill
    \begin{subfigure}[b]{0.45\textwidth}
        \begin{overpic}[width=\textwidth]{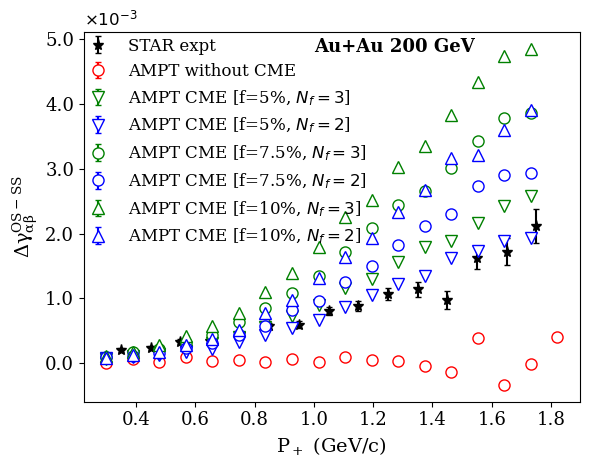}
            \put(20,30){(b)}
        \end{overpic}
    \end{subfigure}
    \caption{ Plot of $\gamma_{\alpha\beta}$ (a) correlation and $\Delta \gamma_{\alpha\beta}$ (b) vs $P_+$  with and without CME for different $f\%$, for 2-flavor and 3-flavor, compared with STAR data in (30-50$\%$) centrality Au+Au collision at $\sqrt{s_{NN}}=200$ GeV}
    \label{fig:gammapp12}
\end{figure}
While the centrality-dependent analysis in Fig. \ref{fig:gammaC1} shows reasonable agreement between model predictions and STAR data, the scale of the gamma correlators exhibits a non-trivial dependence on both the initial charge separation fraction $f$ and the number of quark flavors $N_f$. This dependence becomes more pronounced when examining the $\gamma$ correlators at different transverse momentum ($P_+$) bins at a fixed centrality $30-50\%$ in Au+Au collisions at $\sqrt{s_{NN}}=200$ GeV, as shown in Fig. \ref{fig:gammapp12}. Here, $P_+$ is defined as the average transverse momentum of particle pairs:
\begin{equation}
    P_+ = \frac{P_{T,\alpha} + P_{T,\beta}}{2},
\end{equation}
where $P_{T,\alpha}$ and $P_{T,\beta}$ are the transverse momenta of particles
$\alpha$ and $\beta$, respectively.  Fig. \ref{fig:gammapp12}(a) displays $\gamma_{SS}$ (filled symbols) and $\gamma_{OS}$ (open symbols) as functions of $P_+$, with colors denoting $N_f$ (2 or 3 flavors) and symbols representing $f$ (5\%, 7.5\%, 10\%). The smaller error bars (compared to Fig. \ref{fig:gammaC1}) reflect higher statistics.
At low $P_+$($<1.2$ GeV/c), the 3-flavor scenario with $f=10\%$ (brown diamonds) agrees well with STAR data for $\gamma_{SS}$, while the 2-flavor equivalent overpredicts it. For $P_+>1.2$ GeV/c, the 2-flavor model with $f=10\%$ or the 3-flavor model with $f=7.5\%$ provides better agreement.
The ratio plot (bottom panel of Fig. \ref{fig:gammapp12}(a)) reveals that the 3-flavor, $f=10\%$ case (brown line) stays close to unity for $0.5<P+<1.40$ GeV/c, but other configurations ($f=7.5\%$, $N_f=3$) dominate at higher $P_+$.
This $P_+$-dependent agreement hints at a mass-scale effect or flavor hiararchy. The strange quarks ($N_f=3$) may suppress high-$P_+$ charge separation due to their larger mass and delayed thermalization, while up/down quarks ($N_f=2$) dominate at higher momenta.

Notably, AMPT without initial charge separation (open red squares) reproduces the experimental $\gamma_{OS}$ more accurately than CME-inclusive scenarios, which systematically overpredict the data. This suggests that charge-independent backgrounds (e.g., resonance decays, jet fragmentation) may dominate $\gamma_{OS}$, masking the CME signal. The overprediction of $\gamma_{OS}$ in CME-inclusive models underscores the need for precise background subtraction. Tools like event-shape engineering via $R_{\Psi_2}$ or elliptic flow-scaling methods could help disentangle these contributions.

Fig. \ref{fig:gammapp12}(b) shows $\Delta \gamma$, which isolates CME-driven charge separation. The best match to STAR data occurs for $f=5\%$ and $N_f=2$ (blue inverted triangles), while other initial conditions overpredict $\Delta \gamma$. This implies that smaller charge separation fractions may better reconcile with experimental constraints when combined with 2-flavor dynamics.

\begin{figure}
    \centering
    \begin{subfigure}[b]{0.45\textwidth}
        \begin{overpic}[width=\textwidth]{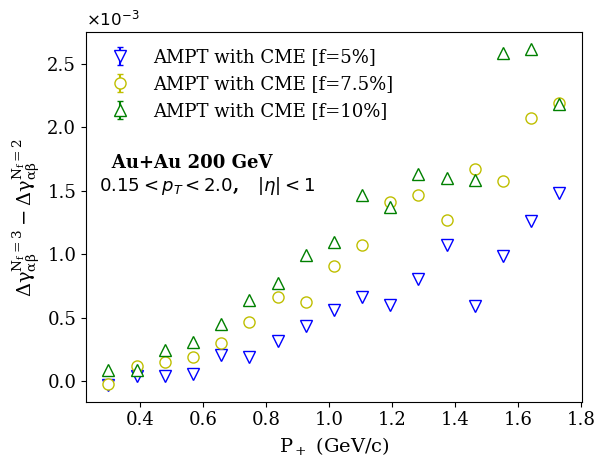}
            \put(52,12){(a)} 
        \end{overpic}
    \end{subfigure}
    \hfill
    \begin{subfigure}[b]{0.45\textwidth}
        \begin{overpic}[width=\textwidth]{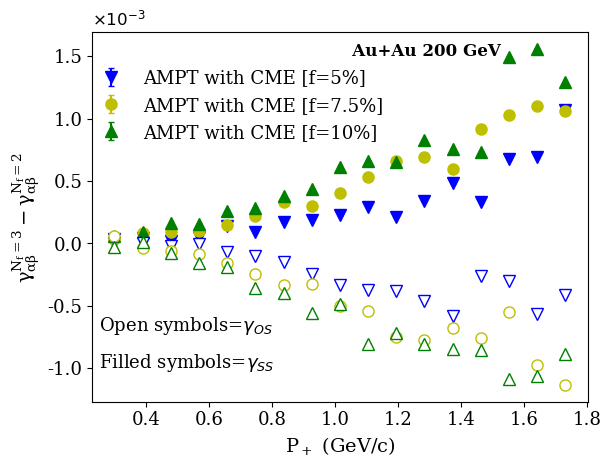}
            \put(52,12){(b)}
        \end{overpic}
    \end{subfigure}
    \caption{Difference between $\Delta\gamma_{\alpha\beta} $ for 2-flavor and 3-flavor CME  with $P_+$ (a) and  $\gamma_{\alpha\beta}$ correlation for 2-flavor and 3-flavor CME with $P_+$(b) for different $f\%$ in Au+Au collision at $\sqrt{s_{NN}}=200$ GeV ($30-50\%$) centrality.}
    \label{fig:gammapp34}
\end{figure}
To further investigate the distinction between the 2-flavor and the 3-flavor CME scenarios, Fig. \ref{fig:gammapp34} presents the  $P_+$-dependence of (a) $\Delta\gamma$ and the gamma correlator $\gamma_{\alpha\beta}$ , for Au+Au collisions at $\sqrt{s_{NN}}$=200 GeV. The plots compare model predictions for different initial charge separation fractions  ($f=5\%, 7.5\%, 10\%$) with $P_+$ defined as the average transverse momentum of particle pairs.
While the absolute difference between 2- and 3-flavor $\Delta\gamma$ in  Fig. \ref{fig:gammapp34}(a) remains small ($\approx 10^{-4}$), the 3-flavor case consistently produces a stronger signal across all $f$ values. This aligns with expectations from anomalous transport, where additional strange quarks enhance parity violation.
The $P_+$ -slope of $\Delta\gamma$ steepens for larger $f$. This could imply a nonlinear response of the QGP to initial axial charge.
The difference between 2-flavor and 3-flavor CME predictions, in Fig.  \ref{fig:gammapp34}(b), is small but systematic, growing modestly with $P_+$. The difference in $\gamma_{\alpha\beta}$ from all initial conditions, exhibit a rise with $P_+$, suggesting a mass-dependent effect in charge separation. Also, the effects of magnetic fields becomes very crucial here.\\

All these suggests that $\gamma$ correlators alone cannot definitively distinguish between 2- and 3-flavor CME scenarios, their $P_+$-dependence reveals a nuanced interplay of quark flavor dynamics and background processes. The sensitivity of $\gamma_{\alpha\beta}$ and $\Delta \gamma$ to $f$ and $N_f$ highlights the challenge of constraining the initial axial charge distribution. In sec \ref{sec:model}, we have used a neural network technique to distinguish the flavor dependencies on the HIC observables.

\section{Flavor dependency in $R_{\Psi_2}$ correlators}
As mentioned in the introduction, the flavor dependency of the $\gamma$ correlator has been studied previously by Huang et. al \cite{Huang1} in Pb+Pb collisions at $\sqrt{s_{NN}}=2.76$ TeV. They have shown that the data from the ALICE 
experiment \cite{ALICE,onderwaater} suggests that the value of the $\gamma$ correlator is different for two flavor and three flavor CME. We have proceeded to calculate the $\gamma$ correlator and obtained a similar behavior in Au+Au collisions at $\sqrt{s_{NN}}=200$ GeV. Now, we present the first theoretical calculation of the $R_{\psi_2}(\Delta S)$ correlator for both 2-flavor and 3-flavor Chiral Magnetic Effect scenarios in Au+Au collisions. This correlator, first proposed by Magdy et. al \cite{magdy1} is defined by the ratio of the charge separation parallel to the magnetic field to the charge separation perpendicular to the magnetic field. If the CME is present, it will give the charge separation only along the magnetic field so we should be getting a concave shaped distribution whose width and minimum will depend on the co-efficients of the P -odd Fourier sine terms. To obtain this correlator, we have used the particle mixing method similar to ref. \cite{mixing,mixing1,magdy2}. We first define the charge separation relative to 
the $\psi_2$ plane in each event as,  
\begin{equation}
\Delta S =  \langle S_{p^+} \rangle - \langle S_{n^-} \rangle
\end{equation}
If $\phi$ is the azimuthal angle of a particle and $\psi_2$ is the $2^{nd}$
order event plane then 
\begin{equation}
\langle S_{p^+} \rangle = \frac{1}{N_p}{\sum\limits_{N_p}{\sin{({\phi}^+_p} - \psi_{2})}}, 
\end{equation}
\begin{equation}
\langle S_{n^+} \rangle = \frac{1}{N_n}{\sum\limits_{N_n}{\sin{({\phi}^-_n} - \psi_{2})}} 
\end{equation}
Here the  + and - signs indicate
the particles’ charges, and $N_p$ and $N_n$ represent the total number of positive and negative charged particles in each events, respectively. If we ignore the charges of the particles, we can come up with the mixed particle correlation in a similar way, 
\begin{equation}
\Delta S_{mix} =  \langle S_{p_{mix}} \rangle - \langle S_{n_{mix}} \rangle
\end{equation}
where we use superscript “mix” to indicate that the particles are of mixed charges.
The correlation functions $C_{\psi_{2}}(\Delta S)$ used to quantify charge separation parallel to the $\vec{B}$ field, is then constructed from the ratio of two distributions: 
\begin{equation}
C_{\psi_{2}}(\Delta S) = \frac{N(\Delta S)}{N(\Delta S_{mix})}
\end{equation}
where  N($\Delta S$) and N($\Delta S_{mix}$) are the distribution over events, of charge separation relative to the $\psi_2$ planes in each event.

Since we need to calculate the $ R_{\psi_2} (\Delta S)$  to distinguish between the background signal and the CME, we now obtain a similar correlation by shifting the $\psi_2$ to $\psi_2$ + $ \pi/2$ and do the same calculations for the new $\psi_2$ + $ \pi/2$ axis. So we obtain the perpendicular correlation as, 
\begin{equation}
C^\perp_{\psi_{2}}(\Delta S) = \frac{N(\Delta S^\perp)}{N(\Delta S^\perp _{mix})}
\end{equation}

\begin{figure}[h!]
	\centering
	\includegraphics[width=0.5\textwidth]{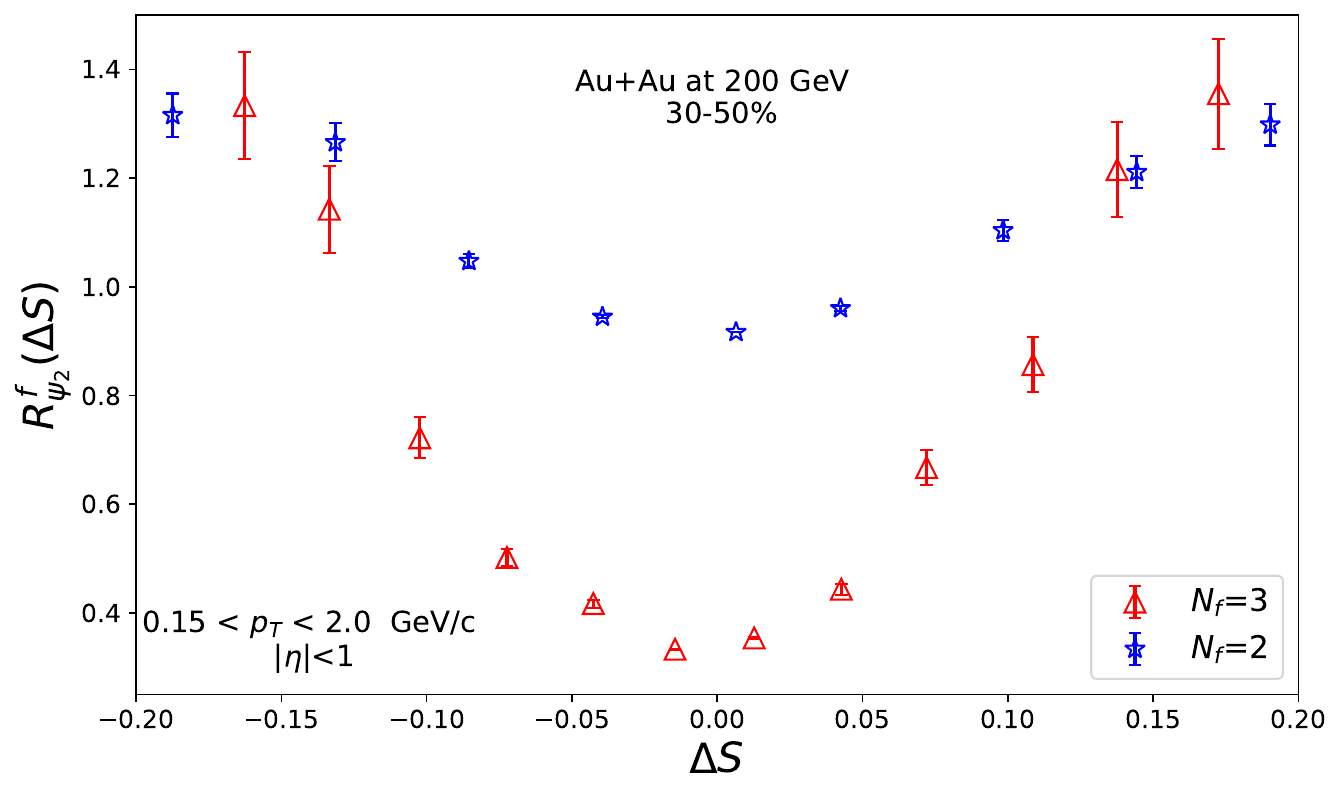}
	\caption{$R^{f}_{\psi_2} (\Delta S)$ in Au+Au collisions at $\sqrt{s_{NN}}=200$ GeV $(30–50\%)$ centrality with  2-flavor(blue) and 3-flavor (red)CME.}
	\label{fig:Rpsiau}
\end{figure}

We then calculate the other CME correlator for 2 -flavor and 3-flavor cases respectively from all these equations and the results are given in Fig. \ref{fig:Rpsiau}.  
\begin{equation}
R^{f}_{\psi_2} (\Delta S) = \frac{C_{\psi_2}(\Delta S)}{C_{\psi_2}^{\perp}(\Delta S)}.
\end{equation}
 As is clearly seen in Fig. \ref{fig:Rpsiau}, the width of the concave distribution is different for 2 flavor and 3 flavor CME.    

The results for $R^{f}_{\psi_2} (\Delta S)$ are consistent with the results obtained from the $\gamma$ -correlator. A narrower shape of the concave distribution implies larger charge separation along the magnetic field. It is to be recalled that for the $\gamma$ -correlator too, the 3 flavor CME shows a larger charge separation as compared to the 2 flavor CME. However, in the case of the $R^{f}_{\psi_2} (\Delta S)$ correlator the 2 flavor and the 3 flavor results are well separated. Fig. \ref{fig:Rpsiau}  therefore appears that the flavor dependency of CME is captured better by this correlator as compared to the $\gamma$-correlator. In order to justify this statement we need to study this correlator across different energy ranges and nuclei species of RHIC and LHC experiments in our future studies.

\section{Classification model with ML algorithm}\label{sec:model}

As mentioned in the introduction, the calculation of both the correlators is computationally challenging. Experimental measurements in collider experiments give us only the final state data and it is well established that the AMPT closely follows the actual experiments and gives us the final state data which is similar to the experimental data. This means that it would be meaningful to use the output data from the AMPT to train ML models to classify the output data in two groups depending on the number of flavors ($N_f$). We therefore proceed to design a classification model to categorize the final state data into two distinct groups according to the number of flavors using the CNN model. 

\begin{figure}[h!]
\centering
	\begin{subfigure}[b]{0.45\textwidth}
        \includegraphics[height=5.5cm,width=8cm]{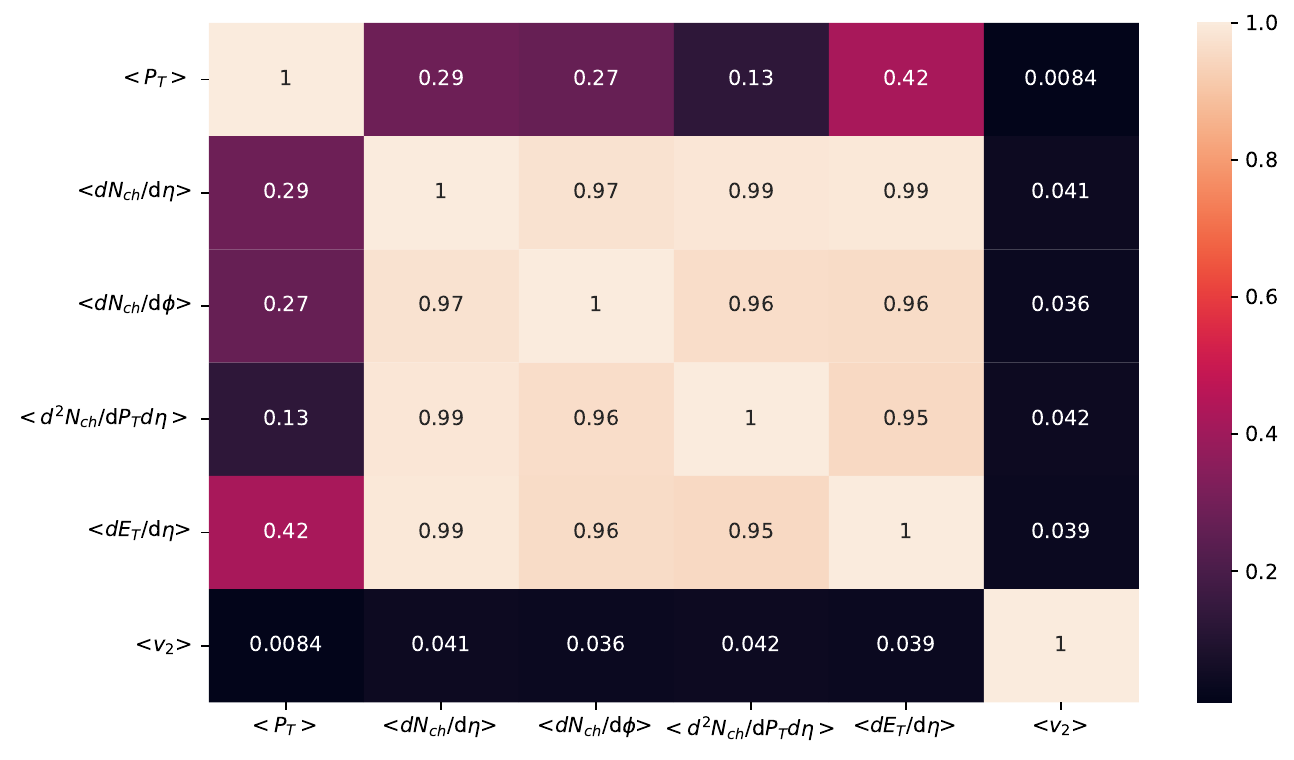}
        \caption*{(a)} 
    \end{subfigure}
    \hfill
    \begin{subfigure}[b]{0.45\textwidth}
        \includegraphics[height=5.5cm,width=8cm]{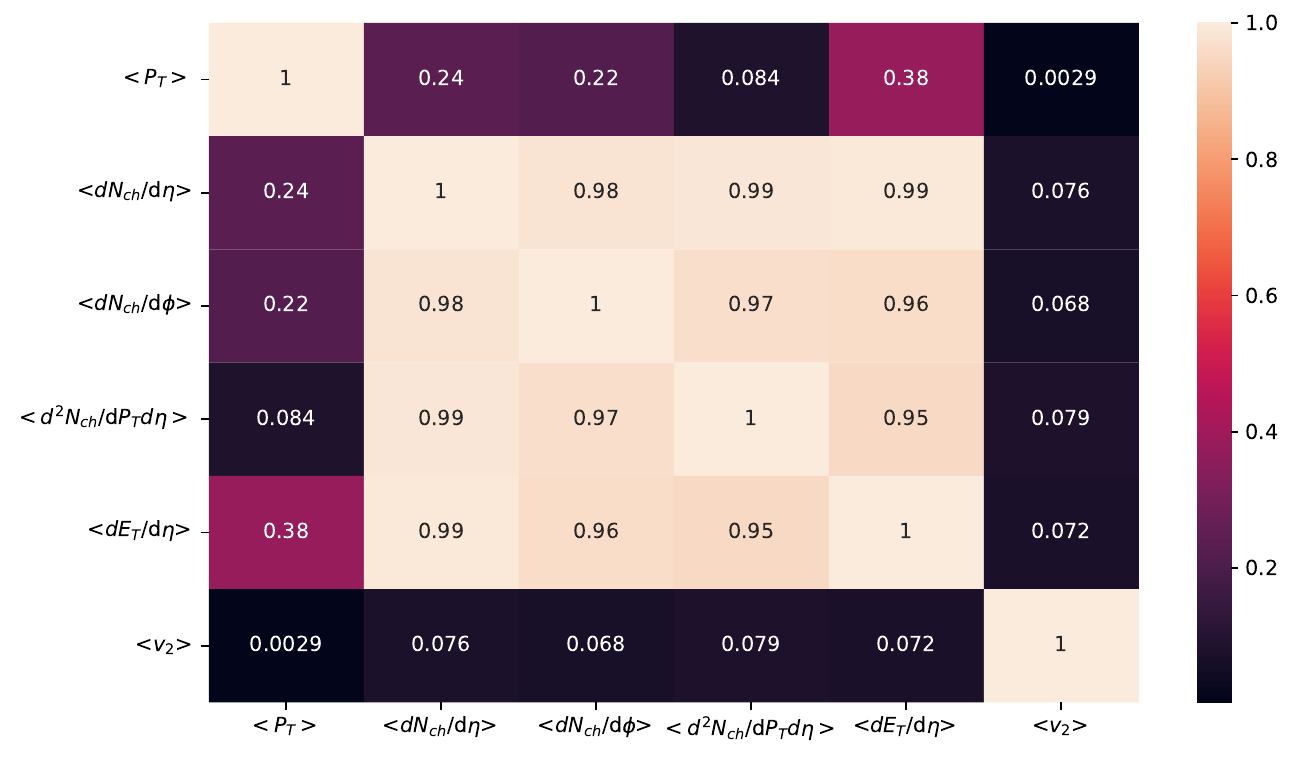}
        \caption*{(b)}
    \end{subfigure}
	\caption{Correlation Matrix for 2-flavor CME(a) and 3-flavor CME(b) in Au+Au collision at (30-50$\%$)centrality at $\sqrt{s_{NN}}=200$ GeV}
\label{fig:cm}
\end{figure}
We first identify the features that we need to use to
categorize the data. Since the CME is related to the number of charged particles, we use the experimental 
observables related to these charged particles as the set of features to design our model. The most 
important amongst these are the charged particle multiplicity at mid-rapidity $(\frac{dN_{ch}}{d\eta})$, the transverse 
momentum $(p_T)$, the elliptic flow coefficient $(v_2)$ and the charged particle multiplicity with respect to the azimuthal angle $(\frac{dN_{ch}}{d\phi})$.
The 3-flavor CME is expected to exhibit broader $p_T$ distributions due to the strange quark contribution, while the 2-flavor dynamics will localize in low-$p_T$ regions. This we have seen in our earlier plots too.
The elliptic flow is also considered here to incorporate the background effects on the CME observable. Other than these we have also taken the 
transverse energy per unit rapidity $(\frac{dE_{T}}{d\eta})$ and the double differential charged 
particle yield $(\frac{dN^2_{ch}}{dp_T d\eta})$  which are also important in determining the collision dynamics of the charged particles.
The chosen pairs probe complementary aspects of the CME dynamics. $\frac{dN_{ch}}{d\eta}$ and $\frac{dE_{T}}{d\eta}$  are sensitive to global charge and energy deposition, which will reflect the axial charge topology while  $\frac{dN_{ch}}{d\phi}$ captures the azimuthal anisotropy, which is potentially correlated with the magnetic field orientation.

The correlation matrix for the classification is shown in Fig.\ref{fig:cm}. It is the standard correlation matrix based on the Pearson’s correlation coefficient. It results in values between $-1$ and $1$. The positive values indicate a pair of 
features that increase or decrease together while the negative values indicate that an increase in one variable implies the decrease in the 
other. The correlation matrix reveals generally weak correlations between observables, with minor differences between the 2-flavor and 3-flavor cases. This aligns with expectations, as CME effects on conventional observables are typically small, particularly when examining flavor-dependent scenarios.  Still, there are several pairs that can be used for the classification of the data. However, the transverse momentum seems to be a feature whose correlation with majority of the other features (except the 
elliptic flow coefficient $(v_2)$) gives a distinct difference 
between the flavor two and flavor three classification. So we choose the pairs selectively having this feature. 

We choose $\frac{dN_{ch}}{d\eta}$ and $p_T$ as one pair. Using this pair we generate two distinct images for the two classes. For each flavor class (2-flavor or 3-flavor CME), we generate 
$5\times10^5$ AMPT events in the ($30-50\%$) centrality range, ensuring sufficient statistics to capture dynamical fluctuations.
 We then calculate the observables on event by event basis. We construct three distinct 2D scatter plots per event class, using correlated final-state observables 
 sensitive to charge separation. The number of events chosen to make each scattered distribution is determined by a sample percentage ($S$).
In our methodology, each image encodes the 2D distribution of events in the space defined by pairs of selected observables. This approach differs from previous neural network applications to CME studies, such as in Ref. \cite{DNN}, where the inputs were constructed from 2D particle spectra binned in transverse momentum ($p_T$) and azimuthal angle $\phi$.
 We generate hundreds of such images for each flavor class. We repeat this process for two more pairs, 
 the $(\frac{dN_{ch}}{d\phi})$, $p_T$; and $(\frac{dE_{T}}{d\eta})$, $p_T$. The images are partitioned into three subgroups: training (70\%), validation (15\%) and testing (15\%). Each image is labeled by its ground-truth CME flavor class (2-flavor or 3-flavor). We used the training subgroup for training our model. The other subgroups were left for testing and validation.\

\begin{figure}[h!]
    \centering
    \includegraphics[width=0.9\linewidth]{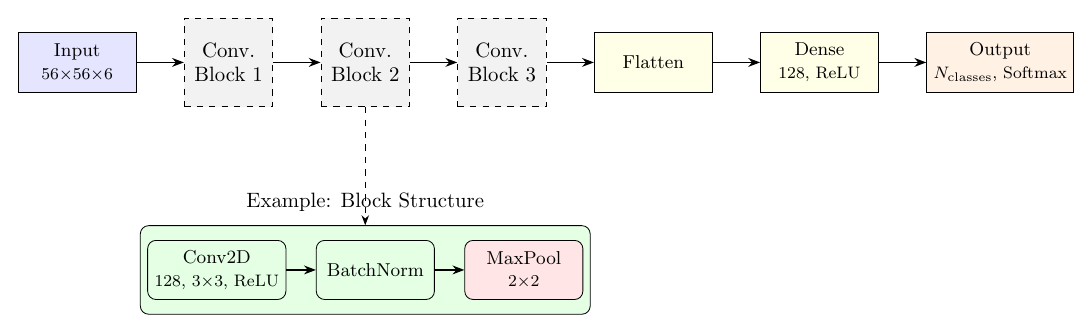}
    \caption{Neural Network architecture to categorize CME flavors}
    \label{fig:enter-label}
\end{figure}
The neural network architecture employed in this study is designed to process input data with dimensions $56\times56\times 6$, which likely represents spatial or feature maps relevant to heavy-ion collision events. Depending on the numbers of input distribution that we want to use for training, the input layer can be single or multi-channel. The network begins with three convolutional blocks (Block 1, Block 2, Block 3). Each block comprises several key components. First is a Conv2D layer with 128 filters of size  $3\times×3$, followed by a ReLU activation function to introduce non-linearity. Then we introduce a Batch Normalization layer to stabilize and accelerate training, followed by a
MaxPooling layer to reduce spatial dimensions and highlight dominant features. The output from the final convolutional block is flattened and a fully connected Dense layer is used for classification. A Softmax activation is used to produce probabilistic class predictions.
The output classifies the data in the two classes. After training the model, we test it with the other subgroup.\\
\begin{figure}[h!]
\centering
	\includegraphics[width=0.5\linewidth]{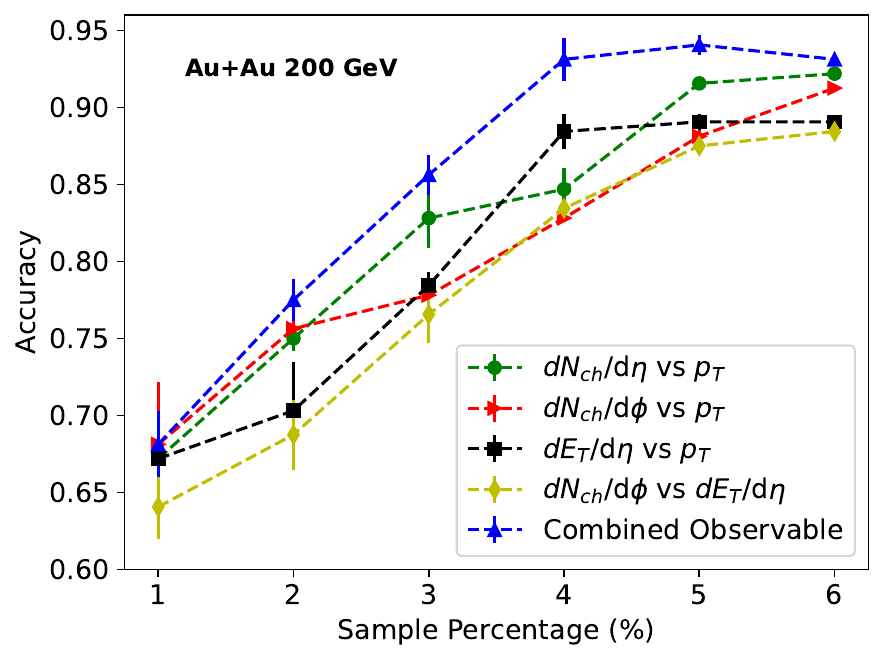}
\caption{Accuracy predictions for the Classification model using individual pairs of features and all pairs of features in Au+Au collision (30-50$\%$) centrality at $\sqrt{s_{NN}}=200$ GeV. The error bars show the accuracy variation across multiple training/testing cycles with randomly selected data subsets.}
\label{fig:acc_au}
\end{figure}

\begin{figure}
\centering
    \begin{subfigure}[b]{0.45\textwidth}
        \begin{overpic}[width=\textwidth]{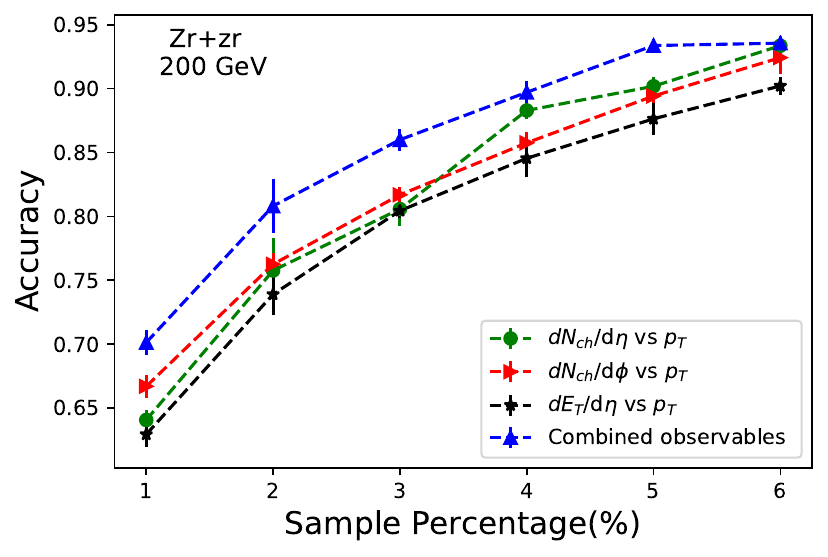}
            \put(45,12){(a)} 
        \end{overpic}
    \end{subfigure}
    \hfill
    \begin{subfigure}[b]{0.45\textwidth}
        \begin{overpic}[width=\textwidth]{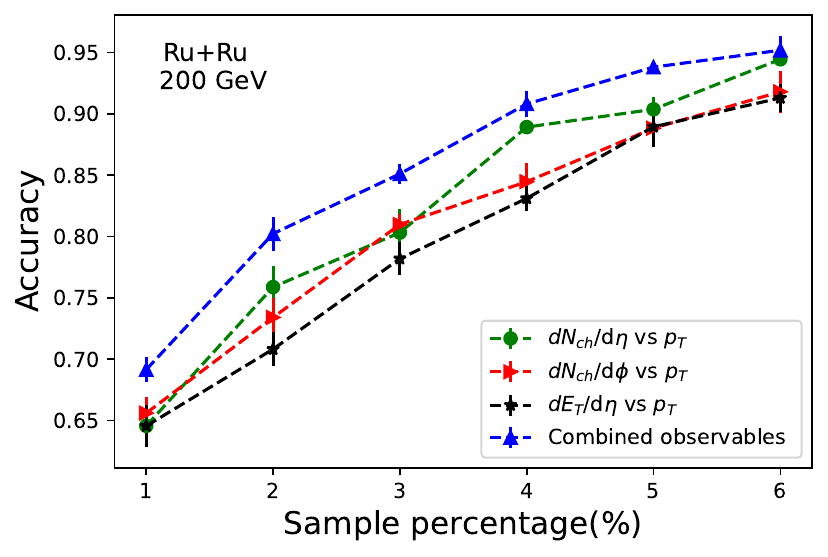}
            \put(45,12){(b)}
        \end{overpic}
    \end{subfigure}
\caption{Accuracy predictions for the Classification model using individual pairs and all pairs of features for Zr+Zr collision (a) and Ru+Ru collision (b) (30-50$\%$) centrality at $\sqrt{s_{NN}}=200$ GeV}
\label{fig:acc_zrru}
\end{figure}

To quantify the performance of our neural network in distinguishing 2-flavor and 3-flavor CME scenarios, we evaluate its prediction accuracy as a function of the sample percentage $S$ for Au+Au, Ru+Ru, and Zr+Zr collisions at  $\sqrt{s_{NN}} = 200$ GeV. We have included the Ru and Zr collisions as recently, there has been a few studies involving the search for CME in these systems too \cite{RuZr}.  This additionally validates the model's applicability to smaller collision systems.
The results are shown in Fig. \ref{fig:acc_au} for Au+Au collisions and Fig. \ref{fig:acc_zrru} for Ru+Ru and Zr+Zr collisions. We used individual pair correlations, as well as a combination of all correlation to check the model efficiency.
The accuracy increases monotonically with $S$, saturating at $\sim 90\%$ for individual observable pairs (e.g., $\frac{dN_{ch}}{d\eta}$  vs $p_T$ , $\frac{dN_{ch}}{d\phi}$ vs $p_T$ ) when $S\ge5\%$. This reflects the neural network’s need for sufficient statistics to resolve subtle flavor-dependent patterns. The error bars represent the distribution of classification accuracies obtained from multiple independent training and testing iterations using different randomly sampled datasets. The highest accuracy ($\sim92\%$) is achieved for $\frac{dN_{ch}}{d\eta}$  vs $p_T$, underscoring $p_T$’s role as a critical discriminator. This aligns with expectations from CME dynamics, where charge separation manifests as a $p_T$-dependent asymmetry.  Integrating all observable pairs (blue curve) boosts accuracy to $95\%$ and reduces the saturation threshold to $S=4\%$, demonstrating synergistic information gain.

A similar classification task is also performed for Zr and Ru collisions and prediction results are shown in Fig. \ref{fig:acc_zrru}. The model maintains robust performance for smaller systems achieving $\sim 90\%$ accuracy at $S=6\%$. For smaller systems, the model requires a higher sample percentage to reach the saturation in accuracy which suggest a slightly diminished discriminative power compared to Au+Au. This may also be a result of shorter lifetimes of small collision systems which contaminates higher non-CME background. The high accuracy confirms that 2- and 3-flavor CME produce distinct imprints in final-state observables, even when these differences are obscured in traditional analyses.\\

\begin{figure}[h!]
\centering
	\includegraphics[height=5cm,width=7.8cm]{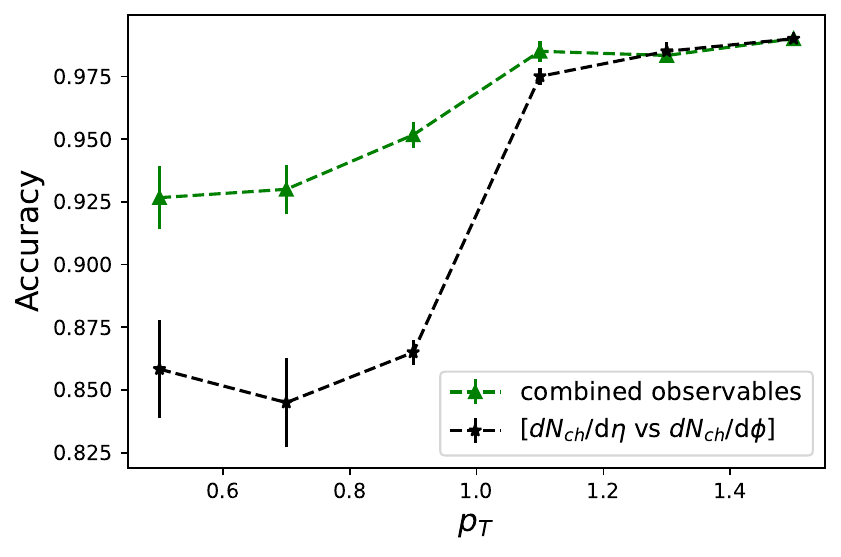}
\caption{Accuracy predictions for the Classification model for different values of transverse momentum in Au+Au collision (30-50$\%$) centrality at $\sqrt{s_{NN}}=200$ GeV. The green and black curves represent the prediction accuracy when the model is trained and tested using the combined set of all observables (green), and individual observable distributions (black).}
\label{fig:acc_pt_au}
\end{figure}

Building on our earlier analysis, we investigate the neural network's ability to distinguish 2-flavor and 3-flavor CME scenarios as a function of transverse momentum. While conventional gamma correlators ($\gamma_{\alpha\beta}$) show increasing flavor separation at higher transverse momentum ($\Delta\gamma^{N_f=3-N_f=2}_{\alpha\beta}$ grows with $p_T$ (Ref. Fig. \ref{fig:gammapp34})), their utility is limited because of several reasons. Statistical uncertainties grow significantly in the large $p_T$ region and
strong background contamination from elliptic-flow-driven effects obscures the magnetic-field-driven CME signal. In the low-pT region, the effect of CME is itself very weak. Thus, traditional correlators cannot distinguish between different flavor scenarios.  Since ML models are known to improve accuracy of prediction in low data regimes, we did a detailed study of the accuracy of the classification model at different values of individual $p_T$.  This means that we trained our model using the data of each individual $p_T$ ranges and tested on a $p_T$ integrated dataset.  For this study too, we evaluated both individual feature pairs and combined observables. The result is shown in Fig. \ref{fig:acc_pt_au}.

As shown in Fig. \ref{fig:acc_pt_au}, the model demonstrates high classification accuracy, consistently exceeding 90\% across all transverse momentum ($p_T$) ranges. Even in the low-$p_T$ region ($<0.8$ GeV/$c$), where traditional methods typically fail, the classifier achieves an accuracy of around 92\%, indicating its robustness and effectiveness in distinguishing between flavor-two and flavor-three CME data. Among the selected feature pairs, those involving charged particle distributions with respect to rapidity and azimuthal angle contribute most significantly to improving classification performance.

\begin{figure}
\centering
    \begin{subfigure}[b]{0.45\textwidth}
        \begin{overpic}[width=\textwidth]{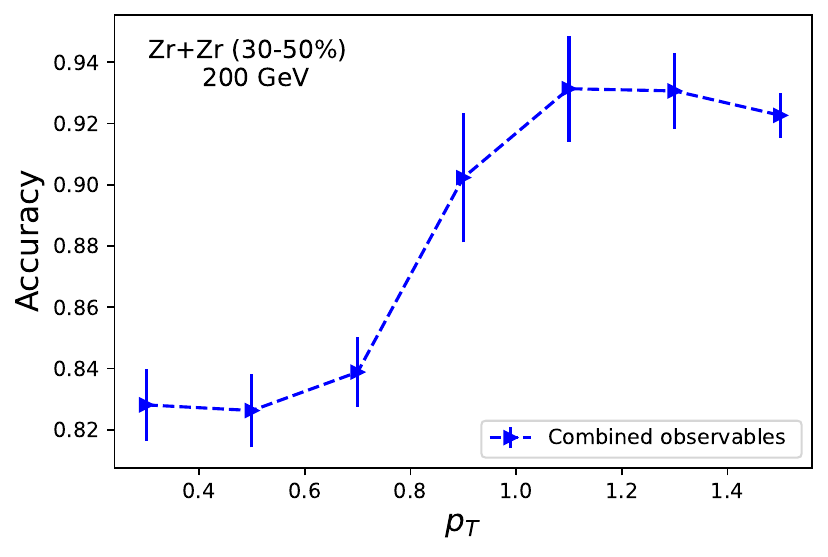}
            \put(50,12){(a)} 
        \end{overpic}
    \end{subfigure}
    \hfill
    \begin{subfigure}[b]{0.45\textwidth}
        \begin{overpic}[width=\textwidth]{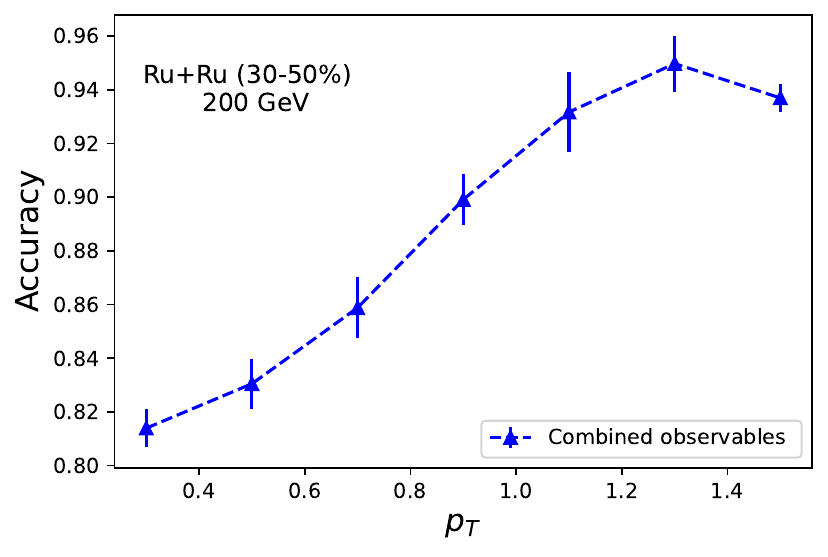}
            \put(50,12){(b)}
        \end{overpic}
    \end{subfigure}
\caption{Accuracy predictions for the Classification model for different values of transverse momentum for Zr+Zr collision (a) and for Ru+Ru collision (b)}
\label{fig:zrru_pt}
\end{figure}

\begin{figure}[h!]
\centering
	\includegraphics[height=5cm,width=7.8cm]{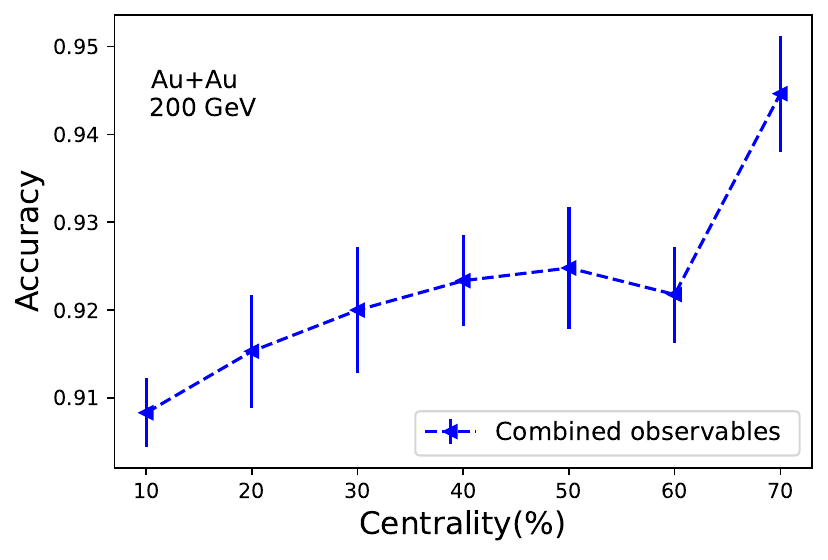}
\caption{Accuracy predictions for the Classification model for different Centrality(\%) values in Au+Au collision.}
\label{fig:Au_cen}
\end{figure} 

In contrast, Fig. \ref{fig:zrru_pt} presents the $p_T$-dependent classification accuracy for Zr+Zr (a) and Ru+Ru (b) collisions, showing clear deviations from the Au+Au results. While the general trend in accuracy versus $p_T$ is similar, the overall accuracy in these lighter systems is noticeably lower, with a decline to approximately 82\% in the low-$p_T$ region. Nevertheless, the model still performs well across the full $p_T$ range, suggesting that flavor-sensitive signatures remain accessible even in smaller collision systems. This underscores the robustness of the neural network approach across different collision geometries and highlights the particular utility of high-$p_T$ observables in these lighter systems.

\begin{figure}[h!]
\centering
    \begin{subfigure}[b]{0.45\textwidth}
        \begin{overpic}[width=\textwidth]{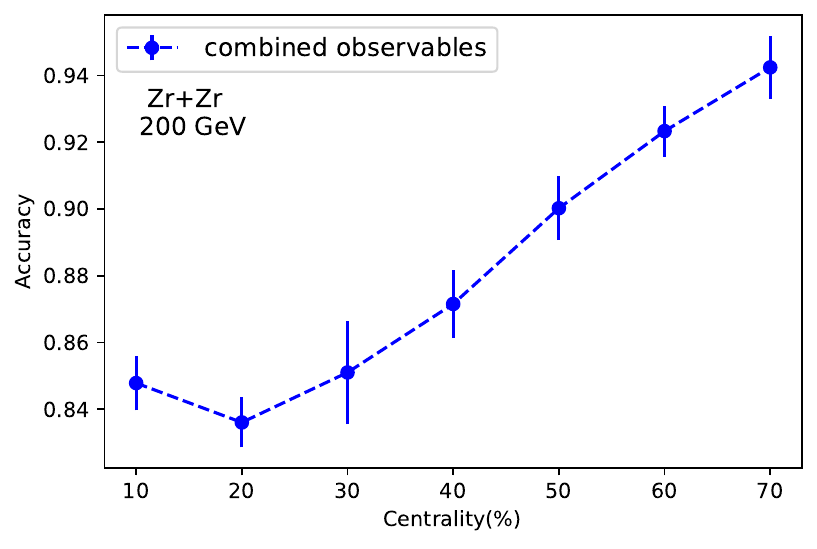}
            \put(50,12){(a)} 
        \end{overpic}
    \end{subfigure}
    \hfill
    \begin{subfigure}[b]{0.45\textwidth}
        \begin{overpic}[width=\textwidth]{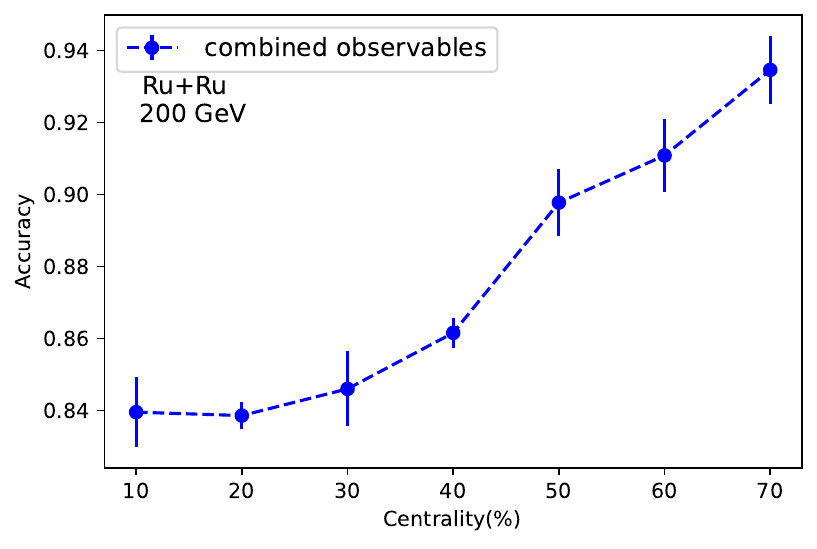}
            \put(50,12){(b)}
        \end{overpic}
    \end{subfigure}
\caption{Accuracy predictions for the Classification model for different centrality($\%$) values  for Zr+Zr collision (a) and for Ru+Ru collision (b)}
\label{fig:zrru_cen}
\end{figure}

To systematically evaluate the neural network’s ability to distinguish 2-flavor and 3-flavor CME scenarios, we extend our analysis across all centrality classes. This investigation is motivated by the centrality-dependent behavior of the gamma correlator observed in Figs. \ref{fig:gammaC1} and \ref{fig:gammac3c4}, where the CME signal diminishes in central collisions due to weaker magnetic fields.
Figure \ref{fig:Au_cen} confirms the model’s high performance in Au+Au collisions, maintaining accuracy above 90\% across all centrality classes, with approximately 91\% accuracy even in the most peripheral (low-centrality) events. Meanwhile, Fig. \ref{fig:zrru_cen} shows the centrality-dependent classification accuracy for Zr+Zr (a) and Ru+Ru (b) collisions. Although the trend with centrality resembles that of Au+Au, the overall accuracy is reduced in these systems, particularly dropping to around 84\% in low-centrality events. This again indicates that while the model remains robust, high-centrality events may provide more reliable observables in smaller systems.

In both low-$p_T$ and low-centrality regimes, the observed reduction in classification accuracy can be attributed to the smaller size of the colliding nuclei—Zr ($A = 40$) and Ru ($A = 44$) versus Au ($A = 197$). The reduced nuclear size leads to a smaller overlap region, shorter system lifetime, and a weaker development of the CME signal, along with a larger relative contribution from non-CME background effects. Additionally, the magnetic fields in these smaller systems are weaker and decay more rapidly, diminishing CME-induced charge separation. Consequently, flavor-dependent patterns become more diffuse in the resulting smaller quark-gluon plasma droplets.

\section{Conclusions} \label{sec:conclusion}
In this work, we have systematically investigated flavor-dependent signatures of the Chiral Magnetic Effect (CME) using a multi-faceted approach combining traditional correlators and machine learning techniques. Typically only the $\gamma$ correlator has been used to study the CME in heavy ion collisions. We have shown that for certain purposes such as the number of flavors involved in the process, it is better to use the  $R_{\psi_2} (\Delta S)$ correlator than the $\gamma$ correlator as the results are more distinct for the $R_{\psi_2} (\Delta S)$ correlator. We have shown that this correlator gives a very clear distinction between two flavor and three flavor CME. In this study, we have introduced the CME into the AMPT model and calculated the $\gamma$ correlator first to confirm that our model is consistent with previous studies. We subsequently use this AMPT to calculate the $R_{\psi_2} (\Delta S)$ correlator for two and three flavor CME. The concave shaped curve has a distinctively different width in the two cases. 

However, it is computationally challenging to obtain this correlator from the data. So we have built a neural network model which can predict the classification based on the output observables. We have used those observables which are related to the charged particle data and built our model from the simulation data that we generated from the AMPT. After rigorous training we find that our model is able to classify the two flavor and three flavor CME to a high accuracy. The analysis also shows the important role of the transverse momentum $p_T$ as a critical discriminator in this classification, which aligns with the fact that the charge separation in CME is a $p_T$ dependent asymmetry.  Our analysis demonstrates that the model successfully discriminates between 2-flavor and 3-flavor CME scenarios across the full transverse momentum spectrum and all centrality classes, using only final-state observables.

This work establishes machine learning as a complementary tool to correlator-based methods, particularly for disentangling multi-flavor effects. The success of our approach suggests that flavor-dependent CME dynamics—particularly the role of strange quarks, can be probed through data-driven methods, bypassing theoretical uncertainties in initial axial charge distributions. The $p_T$-differential performance highlights the importance of momentum-space correlations in quark-level transport, with implications for future CME searches in isobaric collisions or beam-energy scans. Further improvements could integrate event-shape engineering to enhance signal purity or extend the model to other collision systems.

\textbf{Acknowledgment}: 
This work is funded from the IoE research grant No.UH/RITE/PHY/SS/IoE-RC522020/01. S.D is partially supported from the IoE grant.

\end{document}